\documentclass[a4paper,preprintnumbers,11pt]{article}
\usepackage{jcappub, bm, color} 
\usepackage{amssymb,amsfonts,slashed,amsthm,amsmath,graphicx, soul}
\usepackage[caption=false]{subfig}
\newcommand{\eV}{ \ {\rm eV} }

\newcommand{\GeV}{\  {\rm GeV} }

\newcommand{\lmk}{\left(}  
\newcommand{\rmk}{\right)}

\newcommand{\bea}{\begin{array}}
\newcommand{\eea}{\end{array}}
\newcommand{\beq}{\begin{eqnarray}}
\newcommand{\eeq}{\end{eqnarray}}
\newcommand{\eq}[1]{Eq.~(\ref{#1})}

\newcommand{\abs}[1]{\left\vert {#1} \right\vert}

\title{
Isocurvature Perturbations of Dark Energy and Dark Matter
from the Swampland Conjecture
}

\author{
Hiroki Matsui$^{\spadesuit}$
}
\affiliation{$^\spadesuit$ Department of Physics, Tohoku University, 
Sendai, Miyagi 980-8578, Japan}

\author{
Fuminobu Takahashi$^{\spadesuit \diamondsuit \clubsuit}$
}
\affiliation{$^\diamondsuit$ Kavli IPMU (WPI), UTIAS, 
The University of Tokyo, 
Kashiwa, Chiba 277-8583, Japan}
\affiliation{$^\clubsuit$ Department of Physics, Massachusetts Institute of Technology, 
Cambridge, MA 02139 USA}

\author{
and Masaki Yamada$^{\heartsuit}$
}
\affiliation{
$^{\heartsuit}$ Institute of Cosmology, Department of Physics and Astronomy, 
Tufts University, Medford, MA 02155, USA
}

\abstract{
We point out that the recently proposed Swampland conjecture on the potential gradient can lead to  isocurvature perturbations of dark energy, if the quintessence field acquires large quantum fluctuations during high-scale inflation preferred by the conjecture. Also, if the quintessence field is coupled to a dark sector that contains dark matter,  isocurvature perturbation of dark matter is similarly induced. Both isocurvature perturbations can be suppressed if the quintessence potential allows a tracker solution in the early Universe. We find that a vector field of mass $\lesssim$ ${\cal O}(1)$ meV is an excellent dark matter candidate in this context, not only because the right abundance is known to be produced by quantum fluctuations during high-scale inflation without running afoul of isocurvature bounds, but also because its coupling to the quintessence does not spoil the flatness of the potential.
}

\begin{document}

\maketitle
\flushbottom

\section{Introduction
\label{sec:introduction}}

There have been accumulating observational evidence that supports the existence of an inflationary era in the early Universe. The inflation not only solves various initial condition problems of the standard big bang theory but also explains the origin of density 
perturbations~\cite{Starobinsky:1979ty, Kazanas:1980tx, Sato:1980yn, Guth:1980zm,Linde:1981mu,Albrecht:1982wi}. 
Also, the current expansion of the universe is accelerating. In the  $\Lambda$CDM paradigm, both accelerated cosmic expansions are realized by simple slow-roll inflation and the cosmological constant. So far, the $\Lambda$CDM paradigm is in perfect agreement with the current observations such as the Planck data~\cite{Aghanim:2018eyx}.

The great success of the  $\Lambda$CDM paradigm plus slow-roll inflation was recently challenged by the so-called Swampland  conjectures~\cite{Agrawal:2018own}. The Swampland refers to a set of low-energy theories which look consistent from the low-energy perspective but fail to be UV completed with quantum gravity. One of the conjectures sets a lower bound on the potential gradient as
\begin{align}
\frac{\abs{\nabla V}}{V} \ge c, 
\label{SWC}
\end{align}
where $c$ is a numerical constant expected 
to be of order unity, and $V$ is a scalar potential. 
Here and hereafter, we use the reduced 
Planck unit: $8 \pi G = M_{\rm Pl}^{-2} = 1$. 
There are a variety of de Sitter no-go theorem in quantum gravity~\cite{
Gibbons:2003gb,Maldacena:2000mw,Dasgupta:2014pma,
Kutasov:2015eba,
Danielsson:2018ztv,Witten:2001kn}
and the conjecture originates from
a long history of constructing 
de Sitter solutions in string theory.
The cosmological applications and the validity 
of the conjecture have been extensively studied
in Refs.~\cite{Agrawal:2018own,Hertzberg:2007wc,
Andriot:2018wzk,Banerjee:2018qey,Achucarro:2018vey,
Garg:2018reu,Kehagias:2018uem,Lehners:2018vgi,Dias:2018ngv,Colgain:2018wgk,Roupec:2018mbn,Andriot:2018ept,Matsui:2018bsy,Ben-Dayan:2018mhe,Kinney:2018nny,Dasgupta:2018rtp,Marsh:2018kub,Cicoli:2018kdo,Akrami:2018ylq, Heisenberg:2018rdu, Brahma:2018hrd, Choi:2018rze, Das:2018hqy, Dimopoulos:2018upl,Odintsov:2018zai,Ashoorioon:2018sqb,Das:2018rpg,Wang:2018kly,Olguin-Tejo:2018pfq}.

If applied to the inflaton potential, this conjecture leads to a lower bound on one of the slow-roll parameters, $\varepsilon$, as
\begin{align}
 \varepsilon = \frac12 \lmk \frac{\abs{\nabla V}}{V} \rmk^2 \ge \frac{c^2}{2}. 
\end{align}
As long as the quantum fluctuation of the inflaton is responsible for the observed density perturbation, the $\varepsilon$ parameter is related to the Hubble parameter during inflation as~\cite{Aghanim:2018eyx}
\begin{align}
 \varepsilon \simeq 0.010 \lmk \frac{H_{\rm inf}}{10^{14} \GeV} \rmk^2. 
 \label{Hinf}
\end{align}
The current bound on $\varepsilon$ comes from 
non-detection of the primordial gravitational wave.
The Planck $95 \%$ CL limit on the tensor-to-scalar ratio reads,
$r \lesssim 0.10$~\cite{Aghanim:2018eyx}.
Using the slow-roll relation of $r=  16 \varepsilon$, one obtains
\begin{align}
 \varepsilon \lesssim 0.0063. 
\end{align}
Under the Swampland conjecture (\ref{SWC}), this can be expressed as the bound on $c$:
\begin{align}
 c \lesssim 0.11. 
 \label{GW}
\end{align}
So, if $c = {\cal O}(1)$ is taken at a face value, 
there is already a mild tension between the observation and theoretical expectation~\cite{Agrawal:2018own}. In other words,  the Swampland gives a preference to high-scale inflation close to the current bound. Assuming the current bound is saturated, 
the typical Hubble parameter during inflation is $H_{\rm inf} \sim 10^{14}$\,GeV.

Also, the Swampland conjecture  (\ref{SWC}) excludes the cosmological constant as the explanation of the current accelerated expansion, and supports an idea of the quintessence;
the quintessence field $\varphi$ slow-rolls on a potential, and its potential energy drives the accelerated cosmic expansion~\cite{Wetterich:1987fm, Fujii:1982ms, Ford:1987de, Peebles:1987ek, Ratra:1987rm, Fujii:1989qk, Chiba:1997ej, Ferreira:1997au, Ferreira:1997hj, Copeland:1997et, Caldwell:1997ii, Carroll:1998zi, Zlatev:1998tr, Steinhardt:1999nw}. 
When the conjecture is applied to the accelerated expansion of the present Universe, the current bound on $c$ reads~\cite{Agrawal:2018own, Heisenberg:2018yae, Akrami:2018ylq, Heisenberg:2018rdu}
\beq
 c \lesssim 0.6-0.9. 
 \label{constraint2}
\eeq
This leads to the red-shift-dependent lower bound on the equation-of-state parameter
of the dark energy (DE).\footnote{
The equation-of-state parameter $w_0$ of today 
is bounded below as $1+w_0\ge 0.15c^2$~\cite{Agrawal:2018own}
where $w=\rho/p$ for the dark energy perfect fluid.
Currently there is a tension of the Hubble constant $H_0$ 
between the Planck and the local luminosity distance measurements~\cite{riess2018new},
leading to a preference for $1+w<0$~\cite{DiValentino:2016hlg,Qing-Guo:2016ykt,Zhao:2017cud}.
This implies the violation of the null energy condition:
$\rho+p \ge 0$~\cite{ 
Carroll:2003st,Cline:2003gs,Dubovsky:2005xd} and  the 
Big Rip singularity~\cite{Caldwell:2003vq}. The Swampland conjecture strengthens
this tension~\cite{Colgain:2018wgk} and therefore 
it provides another test for the conjecture.
}

The quintessence field may have interactions with other sectors such as dark matter (DM). However, 
its couplings to the standard model are subject to the stringent fifth-force constraints,
which substantially restrict a possible form of the interaction~\cite{Agrawal:2018own, Brennan:2017rbf}. 
In particular, it has been extensively discussed in the literature if the Swampland conjecture is satisfied
before the electroweak and QCD phase transitions~\cite{Denef:2018etk,Murayama:2018lie,Han:2018yrk}. We do not discuss this issue further.\footnote{Note added: after submission of the present paper, the refined conjecture was proposed 
to allow the existence of the local
maxima with a sufficiently large negative curvature~\cite{Ooguri:2018wrx}. } 
 Rather, we
focus on implications of the Swampland conjecture for the inflation and the quintessence as they are the main targets of this conjecture. 
Indeed, one of the striking features of the Swampland conjecture is that it can connect phenomena that took place at vastly different times; both the inflation and the quintessence potentials must satisfy the same simple requirement.

In this Letter we point out that isocurvature perturbation of DE is expected from the conjecture,
if the quintessence field remains light and frozen during inflation. This is because it would then acquire
 large quantum fluctuations during high-scale inflation preferred by the conjecture, which result in the unsuppressed isocurvature perturbation of DE. If the quintessence field is coupled to a dark sector that contains DM, isocurvature perturbation of DM is similarly induced. 
We will also argue that vector DM is a good candidate in this context, 
not only because the right abundance is known to be produced by quantum fluctuations during high-scale inflation without running afoul of isocurvature bounds, but also because its coupling to the quintessence does not spoil the flatness of the potential.
On the other hand, if the quintessence potential allows a tracker-type solution in the early Universe (e.g. in the case of two exponential terms), the fluctuations of the quintessence field will be significantly suppressed as the late-time dynamics is insensitive to the initial condition.

%
\section{Isocurvature perturbations
\label{sec:isocurvature}}

Let us adopt a simple exponential potential for the quintessence field,
\begin{align}
V(\varphi) = \Lambda^4 e^{-c \varphi},
\label{potential}
\end{align}
which saturates the bound (\ref{constraint2}). 
Then the present value of DE should be explained by $V(\varphi)$. 

The dynamics of $\varphi$ with the above potential
has been extensively studied in the literature (see, e.g., Ref.~\cite{Agrawal:2018own}). 
In the early Universe, the quintessence field is almost frozen to the initial value,
if the DE density parameter satisfies $\Omega_{\rm DE} \ll 3/ c^2$.
As the Universe expands, 
the energy density of matter and radiation decreases and that of the quintessence field 
comes to dominate over them. 
Its energy density at present can be fine-tuned by 
adjusting the initial value of $\varphi_*$ or $V(\varphi_*)$. 
The present value of cosmological constant is many orders of magnitude smaller than 
the fundamental scale of physics and requires a fine-tuning in the initial condition 
by a factor of order $10^{-120}$. This is known as the cosmological constant problem, 
which we do not address in this Letter. 

In the above discussion we have assumed that the potential of the
quintessence field does not change its form in the early Universe. 
Here we note that the potential 
is generically 
modified during/after inflation.
For instance, if the quintessence field is coupled to the standard model particles, 
its potential would acquire thermal corrections after reheating. 
Similarly, the quintessence field can acquire a mass of order the Hubble parameter
during inflation if it has a gravitational coupling to the inflaton. 
In fact, in a context of supergravity inflation models, 
 a scalar field generically receives such an effective mass of order the Hubble parameter 
during inflation for a general form of the K\"ahler potential.
This is the so-called $\eta$-problem.
One can avoid the $\eta$-problem only if the scalar mass is protected by
a specific symmetry such as a 
shift-symmetry or the Heisenberg symmetry of no-scale supergravity, or if there is an
accidental cancellation.
Therefore, it depends on how we model the quintessence field in UV theories whether
the mass correction is suppressed during inflation. 
Throughout this paper we simply assume that such corrections to the quintessence potential are negligibly small
and the quintessence remains light during inflation.

It is known that even gravitational couplings to the standard model particles
are tightly constrained by the equivalence principle and fifth force experiments~\cite{Friedman:1991dj, Gradwohl:1992ue, Frieman:1993fv, Gubser:2004uh, Gubser:2004du, Nusser:2004qu}.
If the inflaton and standard model particles are in the same sector, 
it may not be so unrealistic to assume that the Hubble-induced correction during inflation is
also suppressed by some symmetry.  Also, it is non-trivial whether or not the quintessence model works if its potential
is largely deformed during inflation: the quintessence potential may be destabilized or the 
quintessence field might be deviated from the region that leads to the successful explanation of the
current accelerated expansion. Alternatively, if the Hubble correction does not create a new local minimum,
the quintessence field may continue to roll down the (modified) potential until its mass becomes smaller 
than the Hubble parameter where the quantum fluctuation dominates. We will  discuss later a similar phenomenon 
in a case with double exponential terms.

Suppose the quintessence field is extremely light during inflation. Then, it acquires quantum fluctuations,
\begin{align}
\delta \varphi = \frac{H_{\rm inf}}{2\pi}.
\end{align}
This results in the isocurvature perturbation of DE, which is given by 
\begin{align}
\left|\frac{\delta \rho_{\rm DE}}{\rho_{\rm DE}}\right| \simeq  c \frac{H_{\rm inf}}{2\pi},
\label{isocurvature}
\end{align}
well before it comes to dominate the Universe.%
\footnote{
See also Ref.~\cite{Paoletti:2018xet}, where they discussed 
the effect of the isocurvature perturbation of DE in the scalar-tensor theory. 
}
The DE isocurcature perturbation evolves
afterward until present, but it does not change significantly as long as $c$ is less
than unity. 
If the quintessence potential is different from \eq{potential}, 
we expect 
$\abs{\delta \rho_{\rm DE} / \rho_{\rm DE}} \simeq \abs{\nabla V /V} \delta \phi \ge c H_{\rm inf} / (2 \pi)$. 
Hence 
we can regard \eq{isocurvature} as the most conservative estimate
for models satisfying the conjecture (\ref{SWC}).

The DE isocurvature affects the late-time Sachs-Wolfe effects, which appear only
 at large scales of the CMB anisotropies. Therefore, it is subject to the  large uncertainty of the cosmic variance. 
 There is a study on such DE isocurvature perturbations of ultralight axions in a context of the axiverse~\cite{Liu:2010ba, Hlozek:2017zzf}.
According to their analysis, the DE isocurvature bound is much weaker than the DM one.
However, a dedicated analysis is necessary to derive a rigid constraint, and further study is warranted.

\section{DM interacting with quintessence field}

The quintessence field may have couplings to matter fields. 
While its couplings to the standard model
sector are tightly constrained by the fifth force experiments, it is allowed to have sizable couplings to
a dark sector which may contain DM. 
In this Letter, we consider the case where 
the DM mass depends on the
quintessence field, 
\beq
 m_{\rm DM} \propto e^{-c' \varphi}. 
 \label{mass}
\eeq
This kind of model is motivated by the discussion presented in Ref.~\cite{Ooguri:2018wrx}. 
The distance swampland conjecture, which is proposed in Ref.~\cite{Ooguri:2006in}, is related to 
the existence of a tower of massive fields in the hidden sector. According to this conjecture, 
a tower of massive field becomes exponentially light
as a scalar field changes its field value beyond the Planck scale. 
Since the effective field theory breaks down when a massive field becomes lighter than 
an energy scale in consideration, this limits the range of the field value one can change within the effective field theory. 
We can think of the case where DM is made of some of these hidden fields. 
In this case, 
the mass of DM depends on the  quintessence field. This is the motivation of \eq{mass} 
and hence we expect $c' = {\cal O}(1)$. Note that $c'$ is not necessarily equal to $c$
in Eq.~(\ref{SWC}).

\subsection{Model-dependent constraints}
\label{sec:DM1}
Possible interactions between the light quintessence field and DM 
are constrained from particle physics point of view. 
This is because the interactions generically induce quantum corrections to the quintessence potential, 
which may spoil the flatness of the potential. The potential becomes stable against radiative
corrections if one requires that the one-loop corrections to the effective potential 
are smaller than the tree-level one.
Then one obtains~\cite{DAmico:2016jbm}
\beq
 \abs{c'} \lesssim \frac{m_{\varphi}}{\sqrt{G}m_{\rm DM}^2} 
 \simeq c \frac{H_0}{\sqrt{G}m_{\rm DM}^2} 
 \simeq c \lmk \frac{1 \ {\rm meV}}{m_{\rm DM}} \rmk^2. 
 \label{c'bound}
\eeq
Since we expect both $c$ and $c'$ to be of order unity, this requires $m_{\rm DM} \lesssim 1 \ {\rm meV}$. 
This is much smaller than the Tremaine-Gunn phase-space constraint on fermionic DM: $m_{\rm DM} \gtrsim {\cal O}(100 \eV)$~\cite{Tremaine:1979we} (see also Ref.~\cite{Dalcanton:2000hn}) 
and therefore the interacting fermionic DM is inconsistent with this framework.

In the case of bosonic DMs, they can be produced in the early Universe 
with a large occupancy number, 
e.g., via the misalignment mechanism, 
and hence can avoid the phase-space constraint. 
A well-motivated DM model of this kind is an axion-like particle, 
where its tiny mass is generated by a non-perturbative effect 
and a sizable energy density can be produced via the misalignment mechanism. 
However, 
since its mass is quite small, 
it acquires quantum fluctuations during inflation. 
This leads to isocurvature perturbations for DM, 
which is severely constrained by the Planck data as we discussed above. 
Several mechanisms have been proposed to 
suppress its fluctuations by, e.g., making the axion-like particle massive temporarily during inflation~\cite{Linde:1990yj,Linde:1991km,Lyth:1992tw,Kasuya:1996ns,Dine:2004cq,
Folkerts:2013tua,Jeong:2013xta,Higaki:2014ooa,Dine:2014gba,Nakayama:2015pba,Harigaya:2015hha,
Choi:2015zra,Kawasaki:2015lea,Takahashi:2015waa, Kawasaki:2015lpf, Agrawal:2017eqm, Kawasaki:2017xwt}. 
Alternatively, if the Peccei-Quinn symmetry is restored during inflation (or reheating epoch) and 
gets broken after inflation, there is no large correlation for field perturbations beyond the horizon scale. 
This is a simple solution to the isocurvature problem. We note however that 
it may suffer from the domain-wall problem depending on models~\cite{Zeldovich:1974uw, Sikivie:1982qv}. 

An interesting possibility is that the DM is a massive vector field $A_\mu$. 
The longitudinal modes are produced by quantum fluctuations during inflation 
just like a scalar field because of the Goldstone boson equivalence theorem in the relativistic limit. 
Contrarily to the transverse modes, the longitudinal modes do not lead to large anisotropies. 
The equivalence theorem implies that the longitudinal mode multiplied by $m/k$ 
behaves like a scalar field for a large wavenumber $k$, which means that their fluctuations are suppressed at at 
large scales. 
Thus, the vector DM avoids the isocurvature bound because isocurvature perturbations are suppressed at large scales.
The abundance of vector field is calculated as~\cite{Graham:2015rva}
\beq
 \Omega_{\rm DM} h^2 \simeq 0.1 \lmk \frac{m_{\rm DM}}{ 6 \ \mu {\rm eV}} \rmk^{1/2} 
 \lmk \frac{H_{\rm inf}}{10^{14} \GeV} \rmk^2. 
 \label{OmegaA}
\eeq
This is consistent with the fact that the swampland conjecture favors the large-scale inflation (\ref{Hinf}) 
and the constraint of quantum corrections to the quintessence potential (\ref{c'bound}).%
\footnote{
The abundance can be given by \eq{OmegaA} only if its tiny mass is already turned on during inflation. 
This can be realized when the vector field has a Stueckelberg mass. 
However, 
a new conjecture was proposed in Ref.~\cite{Reece:2018zvv}, 
where they show that the Stueckelberg mass cannot be smaller than a certain threshold depending on a cutoff of the theory. 
If we take the cutoff to be larger than $H_{\rm inf}$, 
the Stueckelberg mass need to be less than $0.3 \ {\rm eV}$ according to this new conjecture. }

\subsection{Model-independent constraints}

Now we shall consider constraints that arise independently of the nature of DM. 

The coupled system of DM and
DE is subject to various cosmological observations. 
The quintessence field mediates interactions between DMs, 
whose strength has an upper bound from 
the long-range self-interactions of DM~\cite{Friedman:1991dj, Gradwohl:1992ue, Frieman:1993fv, Gubser:2004uh, Gubser:2004du, Nusser:2004qu}. 
It reads 
\beq
 \abs{c'} \lesssim 0.3. 
 \label{fifthforce}
\eeq
Note that this does not apply to the case where the quintessence 
couples with DM only via spin-dependent interactions, 
e.g. $i e^{\phi} \bar{\psi} \gamma_5 \psi$, where $\psi$ is a fermionic DM.

The DM mass is time-dependent because of the interaction with the quintessence field. 
When $c'$ is positive, 
the mass starts to decrease around the time when the cosmic expansion gets accelerated. 
This is similar to
the decaying DM scenario as the gravitational potential decays, but it is different because the evolution
is correlated with DE. 
The detailed analysis of the system is beyond the scope of this Letter, and further
analysis is warranted. 
Here we simply require that 
the DM mass should not change by more than $10\%$ after the matter dominated epoch.
This is because, in a context of decaying DM models, the DM lifetime was constrained to be
more than ten times longer than the present age of the universe
 by investigating amplitude of matter fluctuations $\sigma_8$~\cite{Audren:2014bca,Enqvist:2015ara}.
The quintessence field varies by a factor of
\beq
 \Delta \varphi = \int^{N_{\rm p}}_{N_{\rm eq}}  \frac{\dot{\varphi}}{H} d N \simeq 0.3 c, 
\eeq
for $c \lesssim 1$ 
after the matter-dominated epoch, 
where $N_{\rm p}$ and $N_{\rm eq}$ represent the e-folding numbers 
at present and 
at the matter-radiation equality, respectively. 
Thus we obtain 
\beq
 c' \lesssim 8/c, 
 \label{constraint-c'}
\eeq
from the constraint on the variation of the energy density of DM 
after the matter-radiation equality. 
It is interesting to note that the product of $c$ and $c'$ is constrained by
the structure formation, which is complementary to (\ref{c'bound}).

If $c'$ is negative, the DM mass starts to increase around the time when the cosmic expansion gets accelerated. 
In this case, the amplitude of matter fluctuations gets enhanced because of deeper gravitational potential of DM. 
Here, we note that the amplitude of matter fluctuation $\sigma_8$ observed by a large scale structure is slightly smaller than that inferred from the CMB measurement~\cite{Aghanim:2018eyx}. The negative $c'$ predicts a larger $\sigma_8$ at present epoch, which would strengthen the discrepancy between the observations of large scale structure and CMB. 
On the other hand, if a subdominant component of DM interacts with the quintessence field,
negative $c'$ has an interesting cosmological scenario (see, e.g., Refs.~\cite{Amendola:2007yx, Amendola:2007yx, Marulli:2011jk, Baldi:2011wy, Wang:2016lxa}).

The isocurvature perturbations of DM are induced by the coupling to the quintessence field 
with the fluctuation $\delta \varphi = H_{\rm inf}/ (2\pi)$ 
as
\begin{align}
 \abs{\frac{\delta \rho_{\rm DM}}{\rho_{\rm DM}}} = \abs{\frac{\delta m_{\rm DM}}{m_{\rm DM}}} =  \abs{c'} \frac{H_{\rm inf}}{2\pi}. 
\end{align}
The Planck observations set a tight bound on the uncorrelated DM isocurvature as~\cite{Akrami:2018odb}
\begin{align}
\beta_{\rm iso} \equiv \frac{{\cal P}_{II}}{{\cal P}_{RR} + {\cal P}_{II}} < 0.038~~~(95\%{\rm \,CL}),
\end{align}
where 
$\beta_{\rm iso}$ is the ratio of the isocurvature power spectrum to the total (adiabatic ${\cal P}_{RR}$ plus isocurvature ${\cal P}_{II}$) power
spectrum. 
This leads to a bound on $c'$:
\begin{align}
\abs{c'} < 1.4 \left(\frac{10^{14} {\rm \, GeV}}{H_{\rm inf}}\right).
\label{result}
\end{align}
Note that this bound on $c'$ is complementary to the one derived by non-observation of the primordial 
gravitational waves (\ref{GW}). For a smaller value of $H_{\rm inf}$, the former becomes weaker while the latter gets tighter. 
These isocurvature perturbations comes from the fluctuation of the quintessence field, 
so that they are independent of the isocurvature modes of the fluctuation of DM itself that are discussed in Sec.~\ref{sec:DM1}.

\begin{figure}[t] 
\centering
\includegraphics[width=8cm]{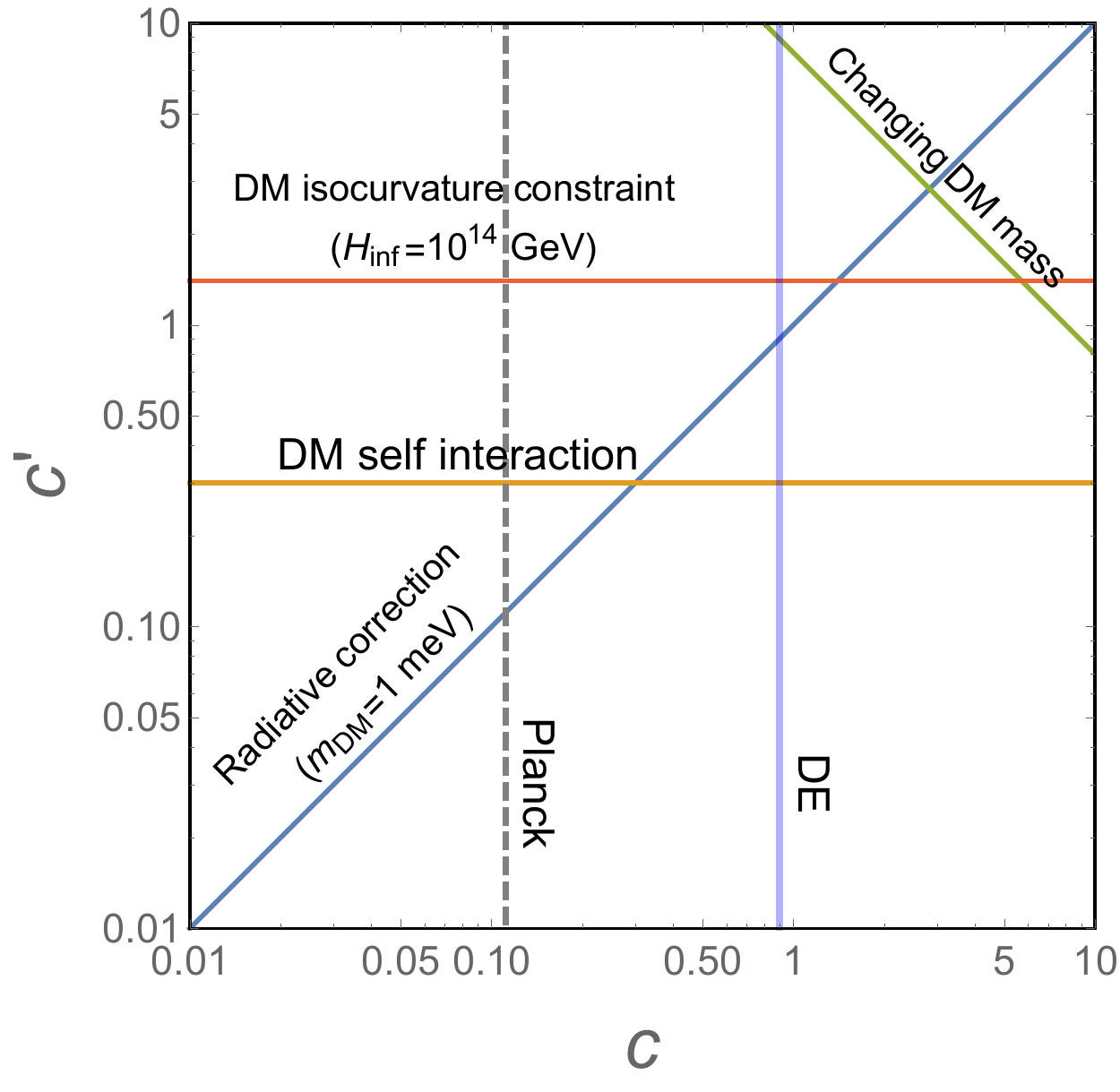} 
\caption{
Constraints on the parameters, $c$ and $c'$,  in the swampland conjecture. See
Eqs.~(\ref{SWC}) and (\ref{mass}) for their definitions. 
The two vertical lines are the upper bound on $c$ and the other ones are the upper bound on $c'$: 
CMB temperature anisotropies by Planck (gray dashed line), 
DE equation of state (light blue line), 
radiative correction (dark blue line), 
DM self interaction (orange line), 
time-dependent DM mass (green line), 
and DM isocurvature constraint (red  line). 
We set $m_{\rm DM} = 1 \ {\rm meV}$ and $H_{\rm inf} = 10^{14} \GeV$ as reference parameters. 
}
\label{fig1}
\end{figure}

We summarize the constraints on $c$ and $c'$ in Fig.~\ref{fig1}. 
The gray line is the upper bound on $c$ from the observation of CMB temperature fluctuations by Planck, 
which is given by \eq{GW}. The light blue line, denoted as DE, is the constraint from the equation of state for dark energy in the quintessence model given by \eq{constraint2}. This is an upper bound on $c$.
The other four lines are the upper bound on $c'$ 
from the radiative correction on quintessence potential (dark blue line, \eq{c'bound}), 
DM self interaction mediated by the quintessence (orange line, \eq{fifthforce}), 
the time-dependent DM mass induced by the dynamics of quintessence (green line, \eq{constraint-c'}), 
and 
DM isocurvature perturabtions generated by the fluctuations of quintessence (red  line, \eq{result}). 
The constraints from radiative correction and DM isocurvature perturbations depend on 
the DM mass and the Hubble parameter during inflation, respectively, 
and we set $m_{\rm DM} = 1 \ {\rm meV}$ and $H_{\rm inf} = 10^{14} \GeV$ as reference parameters.

\section{Discussion and conclusions
\label{sec:discussion}}

The constraints from the isocurvature perturbations can be avoided 
by introducing another exponential term in the quintessence potential with a large coefficient in the exponent, such as $V(\varphi) = \Lambda_1^4 e^{-c_1 \varphi} + \Lambda_2^4 e^{-c_2 \varphi}$~\cite{Barreiro:1999zs}. 
When $c_2 \gg 1$, 
the field $\varphi$ has an attractor solution with the density parameter $\Omega_{\rm DE} = 3/ c_2^2$, 
which removes the dependence of the DE on the initial condition~\cite{Chiba:2012cb}. 
In this case the primordial fluctuation of $\varphi$ does not affect the present value of DE 
and the isocurvature perturbations are absent~\cite{Abramo:2001mv, Kawasaki:2001bq, Kawasaki:2001nx}. 
The DE energy density at the present value is realized by the fine-tuning of parameters in its potential. 

In this Letter we have applied the Swampland conjecture on the potential gradient to both
inflation and quintessence, and studied its implication for isocurvature perturbations of DE and DM. 
In particular, if $c = {\cal O}(1)$ is taken at face value, it gives a preference to 
high-scale inflation which almost saturates the current upper bound on the tensor-to-scalar ratio.
For such high-scale inflation, the quintessence field generically acquire large quantum fluctuations during inflation,
which may results in isocurvature perturbation of DE. However, the current bound on isocurvature perturbation
of DE is  weak and does not give a meaningful bound on $c'$.

We have also studied the case in which the quintessence field is coupled to the mass term
of DM. In this case, CDM isocurvature perturbation is similarly induced. The DM mass should be
smaller than meV to avoid producing too large radiative corrections to the quintessence potential. This 
excludes a possibility of fermionic DM. Among bosonic DM candidates, we find that vector DM with mass $\lesssim$
meV is an excellent candidate, because it is known that the right abundance of massive vector fields 
is naturally produced by quantum fluctuations  during inflation, while isocurvature perturbations
on large scales are suppressed. 
The vector DM has been an active target
of various proposed experiments~\cite{Parker:2013fxa,Chaudhuri:2014dla,Dobrich:2015tpa,Silva-Feaver:2016qhh,
Bloch:2016sjj,Baryakhtar:2018doz}.

%
\section*{Acknowledgments}
F.T.  thanks Mohammad Hossein Namjoo and Alan H. Guth for useful discussion.
F.T. thanks the hospitality of MIT Center for Theoretical Physics and Tufts Institute of Cosmology 
where the present work was done. 
This work is partially supported by JSPS KAKENHI Grant Numbers JP15H05889 (F.T.), 
JP15K21733 (F.T.), JP17H02878 (F.T.), and JP17H02875 (H.M. and F.T.), Leading Young Researcher Overseas
Visit Program at Tohoku University (F.T.), and by World Premier International Research Center
 Initiative (WPI Initiative), MEXT, Japan (F.T.). 

%

\vspace{1cm}
\bibliography{reference}

\end{document}